\documentclass[12pt]{article}
\usepackage{epsf}
\usepackage{a4wide,amssymb,graphicx}
\usepackage{cite}

\newcommand{\beq}{\begin{equation}}
\newcommand{\eeq}[1]{\label{#1} \end{equation}}

\newcommand{\bed}{\begin{displaymath}}
\newcommand{\eed}{\end{displaymath}}
\newcommand{\Ls}{{\Lambda^*}}

\def\bea{\begin{eqnarray}}
\def\eea{\end{eqnarray}}

\begin{document}

\title{$A$ dependence of the 
$\gamma$- and $p$-induced production 
of the $\Lambda(1520)$ from nuclei}

\author{M. Kaskulov$^1$, L. Roca$^2$ and  E. Oset$^1$\\
{\small $^1$~Departamento de F\'{\i}sica Te\'orica and IFIC,
Centro Mixto Universidad de Valencia-CSIC,} \\
{\small Institutos de
Investigaci\'on de Paterna, Aptd. 22085, 46071 Valencia, Spain}\\
{\small $^2$~Departamento de F\'{\i}sica, Universidad de Murcia, E-30100 Murcia,
  Spain} \\
{\tt \footnotesize e-mail:kaskulov@ific.uv.es, luisroca@um.es, oset@ific.uv.es}
}
\date{\today}

\maketitle 

\begin{abstract} 
{\footnotesize
Using results of a recent calculation of the $\Lambda(1520)$ in the nuclear
medium, which show that the medium width is about five times the free width,
we study the A dependence of the $\Lambda(1520)$ production cross section 
in the reactions  $\gamma ~A \to K^+ \Lambda(1520) A^\prime$ and 
$p~ A \to p~ K^+ \Lambda(1520) A^\prime$.  We find a sizable A dependence
in the ratio of the nuclear cross sections for heavy nuclei with respect to a
light one due to the large 
value of the $\Lambda(1520)$ width in the medium, showing that devoted 
experiments, easily within reach in present facilities, can provide good 
information on that magnitude by measuring the cross sections studied here.}
\end{abstract}
\thispagestyle{empty}

\small

\section{Introduction}
The renormalization of particle properties in nuclei is a topic that captures
 permanent attention. Devoted many body calculations are done in order to
 evaluate these properties and parallely there are experimental searches to
 test these theoretical predictions, while other times it goes the other way
 around, with first observation of drastic changes in the nuclear medium.   One
 of the simplest cases from the experimental point of view is the
 renormalization of the $\Delta(1232)$ resonance which can be seen in numerous
 reactions, but perhaps best in the total photonuclear cross section
 \cite{carlos}.  With this  experiment one could test theoretical predictions
 made in \cite{loren} for the $\Delta$ selfenergy and the calculations of the
 photonuclear cross section were done  in \cite{carrasco}. One might guess then
 that other resonances could be tested so easily by means of total
 photoproduction, but the $\Delta$ is maybe an exceptional case where the total
 cross section is absolutely dominated by $\Delta$ excitation in its range of 
 energies.
 This method would obviously not work to test resonances with strangeness where
 other particles with strangeness will have to  be detected in coincidence,
 which will most probably be distorted in the nucleus blurring the signals for
 genuine changes of the resonance in the nucleus.   Other times the cross
 section for production will be a very small fraction of the total cross 
 section
 and pose different but equally difficult problems.  Yet, the determination of
 these properties bears much information on the dynamics of the hadron
 interaction and the efforts to find out these properties are fully justified.

 One of the interesting cases of medium renormalization is that of the
 $\phi$ meson  where theories predict an increase of the width in nuclear
 matter of the order of a factor five to ten \cite{klingl,angelsphi,daniel}.
 Yet, the experimental observation of the change in the width in $\phi$
 photonuclear production is impractical if one looks for a broadening of the
mass distribution of the $\phi$ decay products $(K\bar{K})$ for slow $\phi$'s,
as guessed from \cite{tokiall} and
 further elaborated in \cite{muhlich}.

 One step forward in this direction was given in the experiment
 \cite{ishikawa} where the A dependence of the cross section was studied and
 from there sizable changes of the $\phi$ width in the nuclear medium were
 determined.  Stimulated by this experiment a calculation was done in
 Ref. \cite{adependence} by using the
 theoretical values of this magnitude found in \cite{klingl,angelsphi,daniel}
 which indeed lead to a marked A dependence, although discrepancies of the
 theory
 and experiment remain that require further thoughts.  
 The success of the method
 led to calculations of the A dependence in the proton induced $\phi$
 production in nuclei, showing theoretically that the method is well suited to
 determine changes of the $\phi$ width in the medium \cite{magas}.
 Successively, 
 an experimental proposal to carry out this experiment was approved in the COSY
 facility at Juelich \cite{hartmann}.

     One of the other resonances recently found theoretically with a
spectacular change of the width in the medium is the $\Lambda(1520)$.  This
resonance could qualify as a dynamically generated resonance from the
collaboration of the $\pi \Sigma^*(1385), \bar{K} N $ and $\pi \Sigma$ channels
\cite{sourav,souravnew,Kolomeitsev:2003kt} 
and these channels get largely renormalized in the
nucleus \cite{murat}. One of them is easily visualized:  the coupling of the
$\Lambda(1520)$ to $\pi \Sigma^*(1385)$ is quite large, but the width in vacuum
into this channel is extremely small since there is only phase space  for the
decay through the width of the $\Sigma^*(1385)$.  However, in the nucleus 
the pion
can become a $ph$ and automatically there is plenty of phase space for the
decay. This source alone leads to a width in the medium as large as the free
width. The renormalization of the other channels also leads to large corrections
and finally the total width at normal nuclear matter density turns out to be as
big as five times the free width. Such a spectacular change should be clearly
observable and the purpose of the present work is to present a method of
analysis
based on two reactions, exploiting the A dependence of the cross section.  The
two reactions are the photonuclear excitation and the proton induced production
of the $\Lambda(1520)$.  As we shall see, in both reactions one predicts a
strong A dependence which is amenable of experimental observation in present
experimental facilities.

\section{The $\Lambda(1520)$ in the nuclear medium}
In the description of the $\Lambda(1520)$  properties in the  nuclear medium
we closely follow the formalism developed in Ref.~\cite{murat}.
Here we briefly summarize the main results of that study.

In the nuclear medium the $\Lambda(1520)$ gets renormalized through
the conventional 
$d$-wave decay channels including $\Lambda(1520) \to \bar{K}N$ and
$\Lambda(1520) \to \pi \Sigma$ which 
account for practically all of the $\Lambda(1520)$ free 
width $\Gamma_{free} \simeq 15.6$~MeV~\cite{PDG}. 
In addition in Ref.~\cite{murat}, 
as a novel element, the $s$-wave decay $\Lambda(1520) \to \pi \Sigma^*(1385)$ 
has been considered which is forbidden in the free space 
for the nominal masses of the $\Lambda(1520)$ and $\Sigma^*(1385)$
but opens in the nuclear medium
because of an additional phase space available for the decay products.
The existence of the $\Lambda(1520) \to \pi \Sigma^*(1385)$ 
mode and also the strength of the transition~\cite{souravnew,Sarkar:2005sp} 
is a prediction of the chiral unitary 
models~\cite{Kolomeitsev:2003kt,Sarkar:2004jh}
where 
the $\pi \Sigma^*(1385)$ channel is the most 
important one in the dynamic coupled-channel
formation of the $\Lambda(1520)$ state.

\begin{figure*}[t]
\begin{center}
\includegraphics[clip=true,width=0.7\columnwidth,angle=0.]
{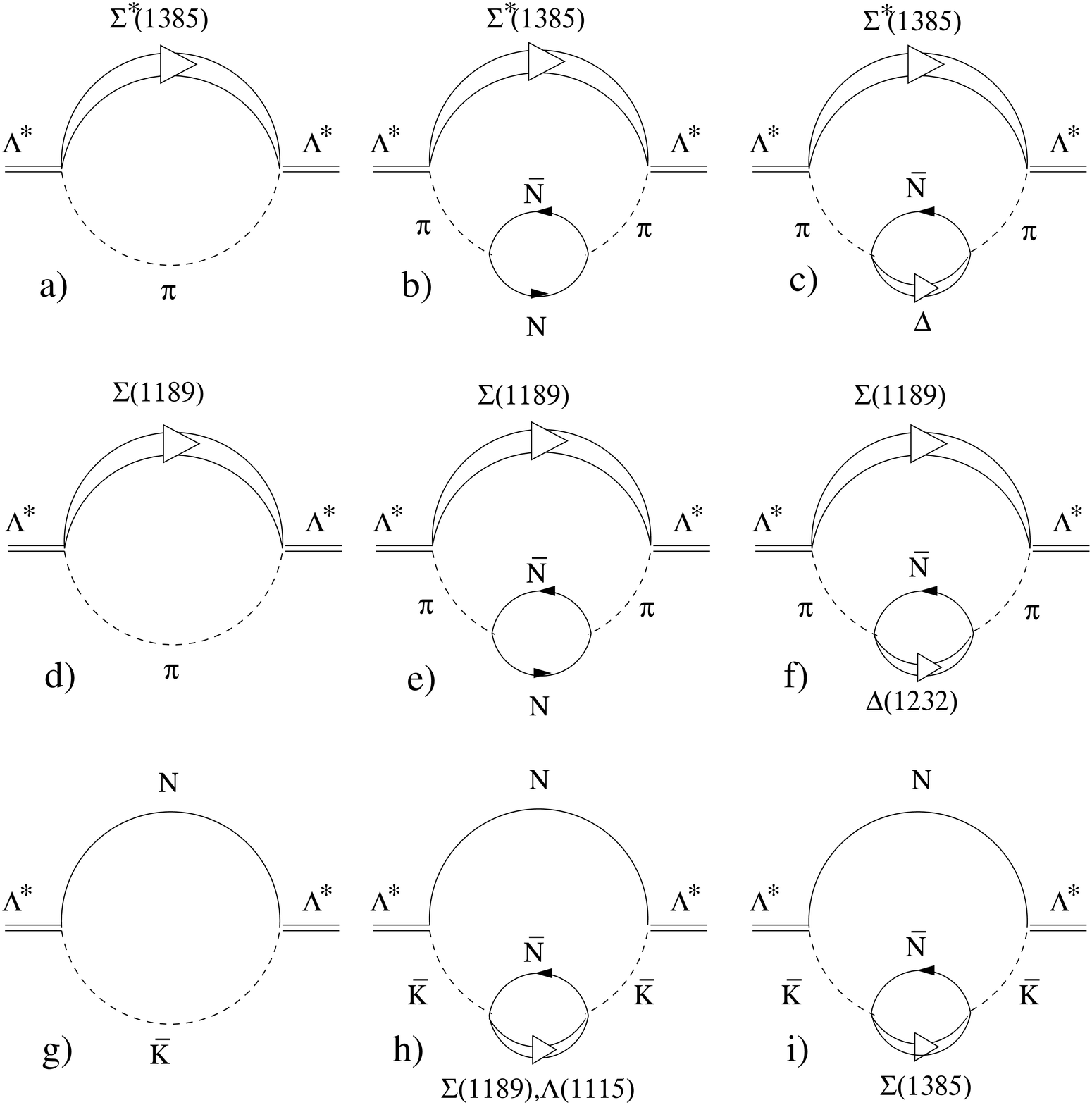}
\caption{\label{F1} \small 
Renormalization of the $\Lambda(1520)$ in the nuclear
medium in the $s$-wave $\pi \Sigma^*(1385)$ and $d$-wave $\bar{K}N$ and 
$\pi \Sigma$ channels.}
\end{center}
\end{figure*}

\begin{figure*}[t]
\begin{center}
\includegraphics[clip=true,width=0.67\columnwidth,angle=0.]
{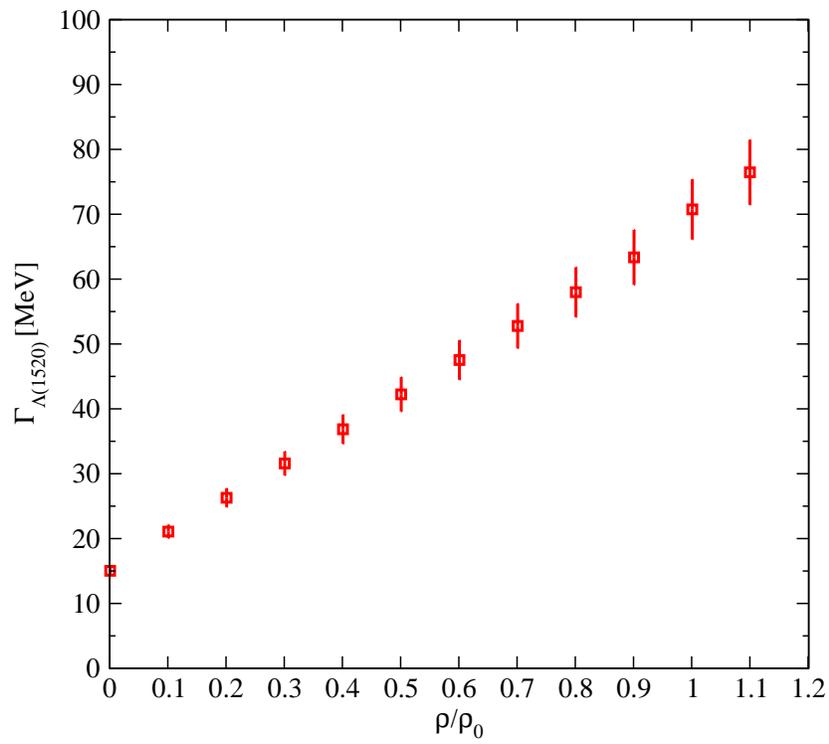}
\caption{ \label{F2} \small
The $\Lambda(1520)$ width in the nuclear medium, from Ref.~\cite{murat}, 
as a function
of the density $\rho/\rho_0$ where $\rho_0$ 
is the normal nuclear matter density. The error bar is explained in the text.}
\end{center}
\end{figure*}

 The model diagrams describing the renormalization
of the $\Lambda(1520)$ in the nuclear medium are shown in Fig.~\ref{F1}.
As one can see, the in-medium propagation of pions in the loops 
is affected by the excitation of the $p-hole$ and $\Delta(1232)-hole$ states 
and in the antikaon $\bar{K}$ case by the excitation of the all relevant 
hyperon-hole states. The intermediate baryons (hyperons) in the loops  
are also dressed  
with respect to their own decay channels properly renormalized in the nuclear 
medium. The later includes the dressing by means of the
phenomenological optical potentials which account for the nuclear binding
corrections, Pauli blocking for the nucleons and
short-range correlations in the $p$-wave transitions induced
by the strong repulsive forces at short inter-baryon 
distances of the Landau-Migdal type~\cite{Kaskulov:2005kr}.
In Fig.~\ref{F2} we show the model prediction 
for the width $\Gamma_{\Lambda^*}$ 
of the $\Lambda(1520)$ at rest and at the nominal pole position
as a function of the nuclear matter density $\rho/\rho_0$ where
$\rho_0=0.17$~fm$^{-3}$ is the normal nuclear matter density. 
The error bars reflect
the theoretical uncertainties due to the choice of the momentum 
cut off in the $d$-wave loops, 
with a cut off constrained by studies of Ref.~\cite{souravnew,Sarkar:2005sp}
(see other details in Ref.~\cite{murat,Kaskulov:2005bp}).
The model predicts a significant change of the
width of the $\Lambda(1520)$ in the nuclear medium
which gets increased by a factor
$\sim 5$ at normal nuclear matter densities.

In the following we address the impact of the in-medium 
width of the $\Lambda(1520)$ 
in the 
$\gamma$- and $p$-induced production of this hyperon from nuclei.
We shall also consider
other 
relevant nuclear effects in the production cross section  which 
will be implemented here by using standard many body techniques,
successfully applied and tested in the past in many works
\cite{carrasco,Salcedo:md} 
to study the interaction of
different particles with nuclei. 
A considerable simplification
will be achieved by 
assuming a local
Fermi sea at each point in the nucleus.  The later provides a very
simple but accurate way to account for the Fermi motion of the
initial nucleon and the Pauli blocking of the final ones.


\section{Nuclear effects in the $\Lambda(1520)$ photoproduction}
We start with the photo-induced production of the $\Lambda(1520)$ in nuclei. 
The
elementary reaction will be
\begin{equation}
\gamma + p \to K^+ + \Lambda(1520)
\end{equation}
where we consider the photoproduction of $K^+ \Lambda(1520)$ pair from protons
only. There are several theoretical works on this 
reaction~\cite{Nam:2005uq,Titov:2005kf,Sibirtsev:2005ns,Roca:2004wt}, 
but the models, as well as the
couplings of the $\Lambda(1520)$ to different channels, are rather different
in all
these works. However, for the purpose of the present work, the detailed
dynamics of the $\gamma p \to K^+ \Lambda(1520)$ reaction is not needed since
we shall evaluate ratios of cross sections 
between different nuclei.

We evaluate the nuclear distortion factor 
due to the $\Lambda(1520)$ absorption
using the eikonal approximation where
the propagation of the $\Lambda(1520)$ in its way out of the nucleus can
be accounted for by means of 
the exponential factor
describing
the probability of loss of flux per unit
length. 

We proceed as follows: let $\Sigma(p_\Ls,\rho(r))$ be
the $\Lambda(1520)$ selfenergy in the nuclear medium,
calculated using
the model of Ref.~\cite{murat}, as a function of its
three momentum, $p_\Ls$, and the nuclear density, $\rho(r)$. We have
for the width

\begin{equation}
\Gamma= - 2{\cal I}m\Sigma \quad ; \qquad
\Gamma\equiv\frac{d{\cal P}}{dt} \ ,
\end{equation}

\noindent
where  ${\cal P}$ is the probability of $\Lambda(1520)$ interaction in the
nucleus, including $\Lambda(1520)$ quasielastic collisions and $\Lambda(1520)$
absorption. We shall not consider the part of the $\mbox{Im} \Sigma$ due to
the quasielastic collisions since, even if the nucleus gets excited, the
$\Lambda(1520)$ will still be there to be observed. Thus, only the absorption
of the $\Lambda(1520)$ is reflected in the loss of $\Lambda(1520)$ events
in the nuclear production. This part of  the $\Lambda(1520)$ selfenergy is the
one calculated in \cite{murat}.
Hence, we have for the probability of loss of
flux per unit length

\begin{equation}
\frac{d{\cal P}}{dl}=\frac{d{\cal P}}{v\,dt}
=\frac{d{\cal P}}{\frac{p_\Ls}{\omega_\Ls}dt}
=-2\omega_\Ls\frac{{\cal I}m\Sigma}{p_\Ls} \ .
\end{equation}

\noindent with  $\omega_{\Ls}=\sqrt{p_{\Ls}^2+M_\Ls^2}$
and the corresponding survival probability is given by

\begin{equation}
\label{EF}
 \exp\left[ -\int_0^{\infty}
dl \frac{(-1)}{p_\Ls}
2\omega_\Ls{\cal I}m\Sigma(p_\Ls,\rho(\vec{r}\, '))\right],
\label{eq:FSI}
\end{equation}

\noindent where
 $\vec{r}\,'=\vec{r}+l\frac{\vec p_\Ls}{|\vec p_\Ls|}$
 with  $\vec{r}$ is the $\Lambda(1520)$ production point inside the
nucleus.  
The study of the $A$ dependence of the total
nuclear cross section due to the $\Lambda(1520)$ absorption,
Eq.~(\ref{eq:FSI}), is the main aim of this work, since it
reflects the modification of the $\Lambda(1520)$ hyperon width in
nuclear matter. 

We proceed further by assuming a local Fermi sea. In this case 
the nuclear cross section which takes into
 account the $\Lambda(1520)$ absorption is given by
\begin{eqnarray}
\label{GammaCS}
{\sigma_{\gamma A}} 
= \frac{M_{\Lambda^*}}{4\pi p_{\gamma}} \int d^3 \vec r
\int \limits_{}^{k_F(r)} 
\frac{d^3\vec{p}_N}{(2\pi)^3}\frac{1}{|\vec P|}
\int \limits_{M_{\Lambda^*}}^{\omega_{\Lambda^*}^{max}}
d\,\omega_{\Lambda^*} 
|\overline{T}|^2
\Theta (1-A^2) \Theta(p_{\gamma} + E(\vec{p}_N) - \omega_{\Lambda^*} - m_K)
\nonumber \\
\times \exp\left[ -\int_0^{\infty}
dl \frac{(-1)}{p_\Ls}
2\omega_\Ls{\cal I}m\Sigma(p_\Ls,\rho(\vec{r} \, '))\right]
\label{F5}
\end{eqnarray}
where the proton density is defined as
$2\int\frac{d^3 \vec{p}_N}{(2\pi)^3}\Theta(k_F(r)-|\vec{p}_N|)=\rho_p(r)$,
with $\vec{p}_N$ is the momentum of protons in the 
Fermi sea, $k_F$ the Fermi
momentum at the local point, $\Theta$ the step function and
$\omega_{\Lambda^*}^{max} = p_{\gamma} + M_N - m_K$.
Also, 
in Eq.~(\ref{GammaCS})
$p_{\gamma}$ the photon momentum in the laboratory frame
(the nucleus is at rest),
$\vec{P} = \vec{p}_{\gamma} + \vec{p}_N$ is the total $\gamma N$ 
three momentum and $A$ is defined as follows
\begin{equation}
A \equiv \frac{1}{2 |\vec{P}| |\vec{p}_{\Lambda^*}|} \left\{
\vec{P}^2 + \vec{p}\,^{2}_{\Lambda^*} + m_{K}^2- 
\Big[p_{\gamma} + E(\vec{p}_N)-\omega_{\Lambda^*} \Big]^2 
\right\}.
\end{equation}
The binding
of the initial nucleon is accounted for by  
$V_s(\vec{r})=-\epsilon_F(\vec{r})= -
k_F^2(r)/(2M_N)$,
and in all places where we have $E(\vec{p}_N)$ we put 
$E(\vec{p}_N) = \sqrt{\vec{p}_N^2+M_N^2} + V_s(\vec{r})$. 

For ${\cal I}m\Sigma$ in the distortion factor we should 
take all sources contributing to the selfenergy 
which do not go into the
final detection channel. Detection of the $\Lambda(1520)$ is done mostly
by reconstruction through its $K^-p$ decay channel, or $\bar{K}^0n$ channel
in present set ups at ELSA.
The $K^- p$ decay channel 
should be
removed
because one can detect it. However, the partial  decay width into the 
$K^-p$  is $\simeq 3.5$~MeV only, out of which, 
some $K^-$ might still be absorbed 
or have quasielastic collisions. This is only a very small fraction
leading to an error of $5\%$. In view of this, and neglecting the small
fraction of the $\Lambda(1520) \to K^-p$ decay followed by $K^-$ absorption in
secondary steps, we consider for ${\cal I}m\Sigma$ in 
Eq.~(\ref{EF}) only the $\Lambda(1520)$ in-medium selfenergy 
calculated in Ref.~\cite{murat} subtracting the free one.
Note also that there is no coherent production here
since there is always conversion of an initial
proton into a $\Lambda(1520)$, i.e.
the nucleus does not remain in its ground state.

We shall evaluate the ratio between the nuclear cross sections in
heavy nuclei and a light one, for instance  $^{12}C$, since in
this way, many other nuclear effects not related to the
distortion of the $\Lambda(1520)$ cancel in the
ratio, as was shown in Ref.~\cite{magas}.
In the ratio of cross sections
we shall eliminate $|\overline{T}|^2$ which should 
cancel if the later 
is not much energy dependent. In this respect, let us consider that the 
$K^+\Lambda(1520)$ production will be measured by looking at 
$d\sigma_{\gamma A}/d M_{K^- p}$ and selecting the contribution of the peak
of the $K^-p$ invariant mass around the $\Lambda(1520)$. This gives a very
restricted phase space where the cancellation of $|\overline{T}|^2$ is 
justified.

In Eq.~(\ref{GammaCS}) we have considered the $K^+\Lambda(1520)$ production on
protons of the target. In principle, we could also have $K^+\Lambda(1520)$
production in two step processes like $\gamma n \to K^0 \Lambda(1520)$ followed
by $K^0 p \to K^+n$. The chances for this two step reaction are not large, but
in any case, one of the good things to make ratios of cross sections on
heavy nuclei to light nuclei is that the effect of these two step processes is
highly reduced in the ratio~\cite{magas}.

\section{Nuclear effects in the $p$-induced $\Lambda(1520)$ production}

In a similar way to what was obtained in Ref.~\cite{magas}
for the $\phi$ meson production in the $p$-induced reaction, we
expect the $A$-dependence of the $pA\to A'K^+\Lambda(1520)$ reaction in
nuclei to provide also a conclusive test of the modification of
the $\Lambda(1520)$ width in the nuclear medium. 
Within the local Fermi sea approach the $pA$-nuclear cross
section can be evaluated, as a first approximation, as:

\begin{equation}
\sigma_{pA}(p_{Lab})=4\int d^3r\int\frac{d^3 k}{(2\pi)^3}
 \Theta(k_F-|\vec{k}|) \sigma_m(p_{Lab},\vec k, \vec r)
\label{eq:sigmaA1}
\end{equation}

\noindent 
where $p_{Lab}$ is the momentum
of the incident proton and $\sigma_m$ the elementary 
$pp\to pK^+\Lambda(1520)$ and
$pn\to
nK^+\Lambda(1520)$ 
average cross section in the nuclear medium,
which will be defined later in Eq.~(\ref{eq:sigmam}).

One of the main differences of the $p$-induced reaction with respect to the
photoproduction case
previously discussed is the fact that the incident proton is strongly distorted
in its way in till the
reaction point. This effect can be properly considered by adding in
Eq.~(\ref{eq:sigmaA1}) the following eikonal factor:

\begin{equation}
\exp\left[-\int_{-\infty}^z\sigma_{pN}(p_{Lab})\rho(\sqrt{b^2+z'^2})dz'\right]
\label{eq:ISI}
\end{equation}

\noindent
where $z$ and $\vec{b}$ are the position in the beam
axis and the impact parameter, respectively, of the production
point $\vec{r}$ of Eq.~(\ref{eq:sigmaA1}). In Eq.~(\ref{eq:ISI})
$\sigma_{pN}$ is the total $pp$ and $pn$ averaged experimental
cross section, taken from \cite{PDG}, for a given incident
proton momentum.  Eq.~(\ref{eq:ISI}) represents the probability
for the proton to reach the reaction point without having a
collision with the nucleons,  since $\sigma_{pN}\rho$ is the
probability of proton collisions per unit length. There is of course the 
possibility of having the reaction through two step collisions, but this was
discussed in \cite{magas} and found to cancel to a great extent in the ratios
of cross sections of heavy to medium nuclei.

The elementary cross section in the nuclear medium for
$p(p_{Lab})+N(k)\to N(p_1)+K^+(p_2)+\Ls(p_\Ls)$ reaction is

\begin{eqnarray}\nonumber
\sigma_{m}(p_{Lab},\vec k,\vec r)&\sim&
\frac{1}{|\vec p_{Lab}|}
\int d\Omega_\Ls \int dp_\Ls p^2_\Ls
\int{dp_1}\frac{p_1}{P}\frac{1}{E(p_1)\omega(p_\Lambda^*)}\times\\
&\times &
\overline{\sum_{s_i}}\sum_{s_f}|T|^2
\Theta(p_1-k_F(r))\Theta(p_2-k_F(r))\Theta(1-A^2) \nonumber \\
&\times & \Theta(E(p_{Lab})+E(k)-E(p_1)-\omega(p_\Ls)-m_K)
\label{eq:sigmam}
\end{eqnarray}
\noindent
up to some global constants irrelevant
 in the final results since
they will cancel  when we will evaluate the ratio of different
nuclei to $^{12}C$. In Eq.~(\ref{eq:sigmam})
$P=p_{Lab}+k-p_\Ls$, and $A$
provides the cosinus of the angle between $\vec P$ and $\vec
p_1$, 
$$A\equiv \frac{1}{2|\vec P||\vec p_1|} \left\{m_K^2+\vec
P^2+\vec
p\,^2_1-[E(p_{Lab})+E(k)-E(p_1)-\omega(p_\Ls)]^2\right\},$$
with $E(q)=\sqrt{M^2+\vec q\,^2}$,
$\omega(q)=\sqrt{m_\Ls^2+\vec q\,^2}$. In Eq.~(\ref{eq:sigmam})
the azimuthal angle of $\vec{p}_1$ with respect to $\vec{P}$
has already been integrated, assuming that $|T|^2$ does not
depend on this angle.  

Gathering all these results, the final expression for the $\Ls$
production cross section in nuclei reads, up to a global constant
factor:

\begin{eqnarray} \label{eq:sigmaA2}\nonumber
\sigma_A(p_{Lab}) &\sim& \frac{1}{|\vec{p}_{Lab}|}
\int d^2b\int_{-\infty}^{\infty}dz
\exp\left\{-\int_{-\infty}^z\sigma_{pN}(p_{Lab})
\rho(\sqrt{b^2+z'^2})dz'\right\} \\ \nonumber
& & \times \int  d^3k \int{dp_1} \int d\Omega_\Ls \int dp_\Ls
\frac{|\vec{p_\Ls}|^2|\vec{p_1}|}{|\vec{P}|E(p_1)\omega(p_\Ls)}
\nonumber \\ 
&& \times \overline{\sum_{s_i}}\sum_{s_f}|T|^2
\exp\left\{ -\int_0^{\infty}
dl \frac{(-1)}{|\vec{p_\Ls}|}
2\omega_\Ls{\cal I}m\Sigma(p_\Ls,\rho(\vec{r}\,'))\right\}
\\
& & \times 
\Theta(k_F-|\vec{k}|)\Theta(p_1-k_F(r))\Theta(p_2-k_F(r))\Theta(1-A^2)
\nonumber \\
& & \times  \Theta(E(p_{Lab})+E(k)-E(p_1)-\omega(p_\Ls)-m_K)
\end{eqnarray}

\section {Results and discussion}
\label{res}


We
performed calculations for the
following nuclei:   ${}^{12}_6C$, ${}^{16}_{8}O$,
${}^{24}_{12}Mg$,  ${}^{27}_{13}Al$, ${}^{28}_{14}Si$,
${}^{31}_{15}P$,   ${}^{32}_{16}S$,  ${}^{40}_{20}Ca$,
${}^{56}_{26}Fe$, ${}^{64}_{29}Cu$,  ${}^{89}_{39}Y$,
${}^{110}_{48}Cd$,  ${}^{152}_{62}Sm$,  ${}^{208}_{82}Pb$,
${}^{238}_{92}U$. 

\begin{figure*}[t]
\begin{center}
\includegraphics[clip=true,width=0.79\columnwidth,angle=0.]
{Figure3a.eps}
\includegraphics[clip=true,width=0.79\columnwidth,angle=0.]
{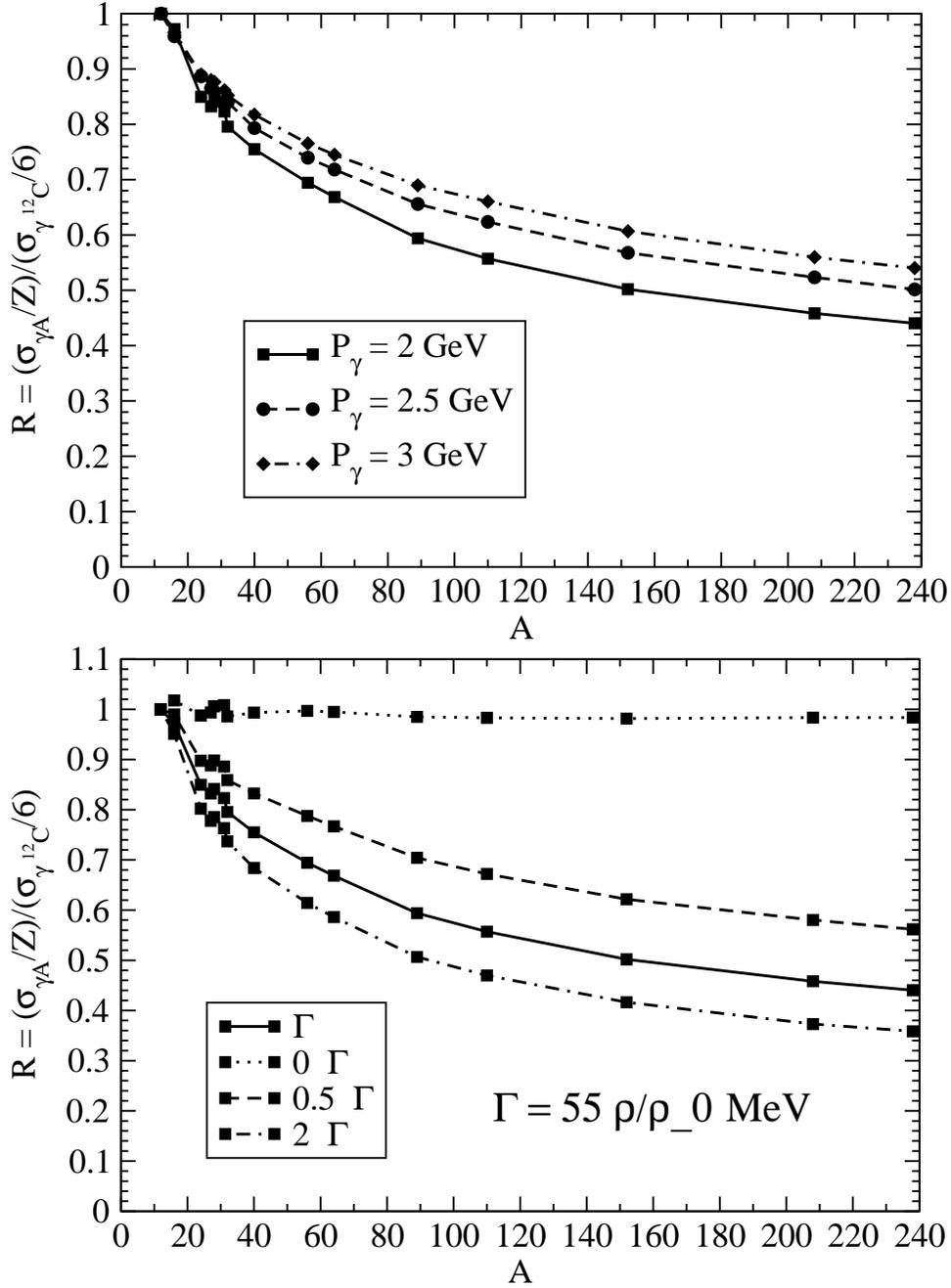}
\caption{\label{F3} \small The nuclear cross section for the 
$\gamma$-induced production of the $\Lambda(1520)$. Top panel:
Ratio of the nuclear cross section normalized to
$^{12}C$ for $p_{\gamma}=2, 2.5$ and $ 3\textrm{ GeV}$.
Bottom panel: The nuclear cross section normalized to
$^{12}C$ for $p_{\gamma}=2\textrm{ GeV}$ and multiplying the 
in-medium width of the $\Lambda(1520)$ by different factors.}
\end{center}
\end{figure*}

In Fig.~\ref{F3}~(top) we present our results for the ratio of the nuclear 
cross sections normalized to
$^{12}C$
as a function of the mass number $A$. The curves 
correspond to the incident $\gamma$ momenta in the laboratory frame
$p_{\gamma} = 2$~(solid),~2.5~(dashed) and $3$~GeV~(dot-dashed).
Recall, that contrary to the $p$-induced production, in the $\gamma$-induced
reaction the elementary cross section is defined with respect 
to the protons
in the Fermi sea only. 
Hence, 
the nuclear cross sections are normalized to the number of protons $Z$
in the corresponding nucleus, $(\sigma_{\gamma A}/Z)$.

As one can see, we obtain a significant reduction 
of flux relative to $^{12}C$ 
which can reach the value $\simeq 0.4\div 0.5$ for heavy nuclei.  
It is also instructive to  renormalize artificially the
model prediction for the $\Lambda(1520)$ selfenergy 
by some factors and look on the corresponding
ratio of the nuclear cross sections. In Fig.~\ref{F3} (bottom panel) 
we show our results for
$f \cdot {\cal I}m\Sigma $ entering the eikonal factor,
 where the factor $f $  is taken
$f=0,\, 0.5,\, 1 \,$ and $2$. The calculations are 
performed for $p_\gamma = 2$~GeV. These results are instructive because they
tell us with which accuracy the ratios must be measured experimentally to
induce a certain value of the $\Lambda(1520)$ in-medium width.

\begin{figure*}[t]
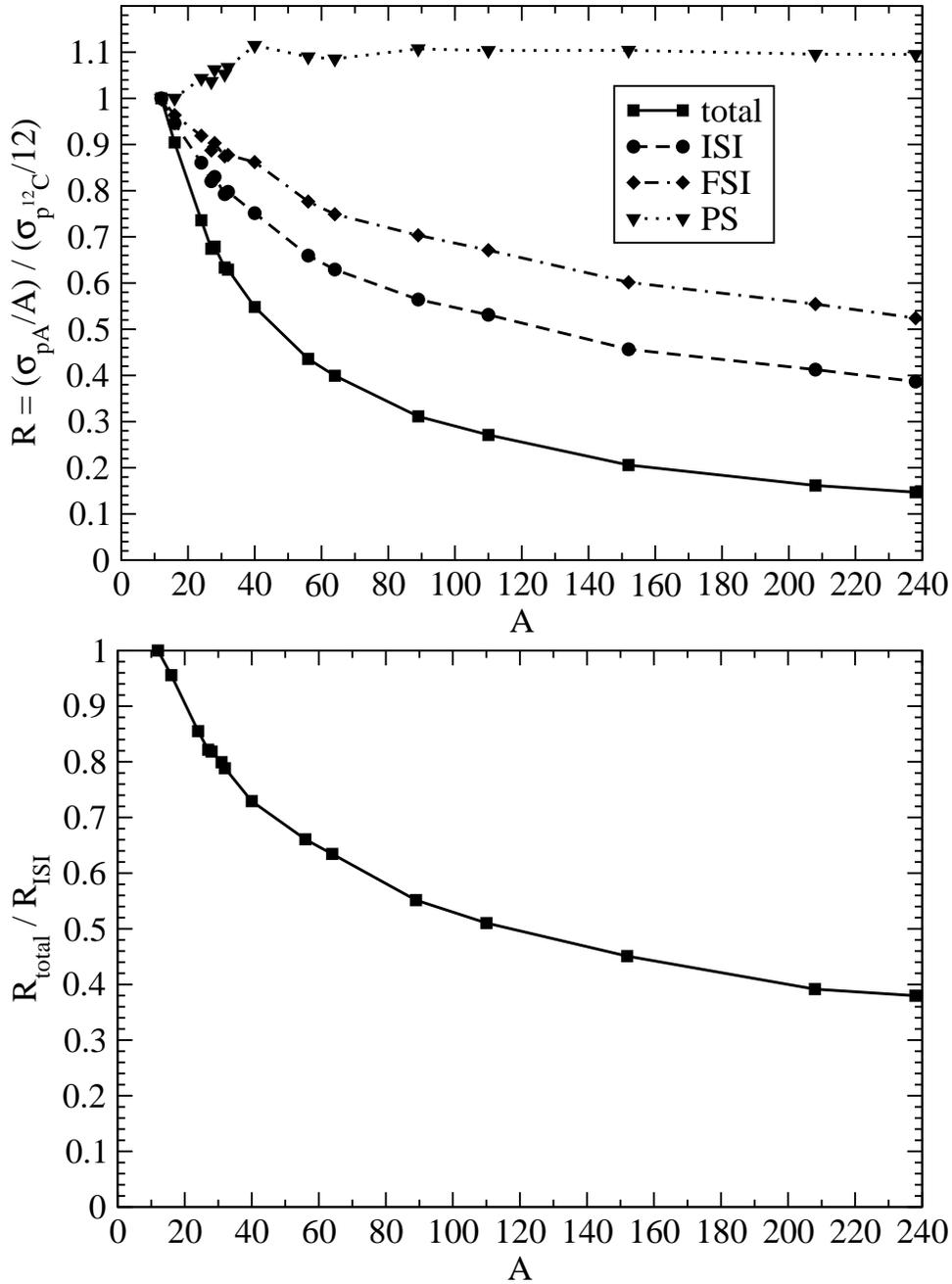

\begin{center}
\includegraphics[clip=true,width=0.79\columnwidth,angle=0.]
{Figure4a.eps}
\includegraphics[clip=true,width=0.79\columnwidth,angle=0.]
{Figure4b.eps}
\caption{\label{F4} \small The nuclear cross section for the 
$p$-induced production of the $\Lambda(1520)$.
Top panel: 
The nuclear cross section as a function
of the mass number $A$
normalized to $^{12}C$
 for proton kinetic energy
$T_p=2.9\textrm{ GeV}$.
 Different lines correspond to separate contributions from the
phase-space (PS), initial state interaction (ISI), final state interaction 
(FSI)
and total cross section.
Bottom panel: Ratio of the total result to ISI for
$T_p=2.9\textrm{ GeV}$.}
\end{center}
\end{figure*}

Next we present the results for the $p$-induced  production. 
In Fig.~\ref{F4}~(top) we show $(\sigma_A/A)$
normalized to the value for $^{12}C$ as a function of the
mass number and for a projectile energy of
$T_p=2.9\textrm{ GeV}$.
The different lines correspond to separate contributions
from the phase space (PS), Eq.~(\ref{eq:sigmaA1}); phase space
and initial state interaction, Eq.~(\ref{eq:sigmaA1}) including
the distortion factor of Eq.~(\ref{eq:ISI}), (ISI);  phase
space and final state interaction, Eq.~(\ref{eq:sigmaA1})
including the distortion factor of Eq.~(\ref{eq:FSI}), (FSI);
and complete calculation, (total), i.e. the simultaneous
contribution of all the effects, Eq.~(\ref{eq:sigmaA2}).

As we can see in the figure, the (PS) curve is quite stable with
respect to $A$, almost saturating from about $A\sim 50$ on. This feature
is common for both, $p$-induced and $\gamma$-induced productions.
This is just a geometric effect of the density profiles of the
different nuclei which makes that the average density of each
nuclei almost saturates with $A$.
 By looking at the dot-dashed line, the
effect of the distortion of the incident proton in its way 
through the nucleus is significative. This is the most important
difference of the production induced by protons with respect to
the photoproduction case. The dashed line represents the effect of
only considering the FSI of the $\Ls$, but not the ISI. The
difference of this curve to the PS curve is only due to the
modification of the $\Lambda(1520)$ in the medium. 
 The solid line
represents the calculation obtained including all the effects
considered.
Now if we look at ISI and FSI curves, we see that in both cases
there is a sizeable decrease of the observable,
 particularly for ISI, which
shows a stronger $A$ dependence.
Although the ISI and "total" curves are almost parallel,
the absolute values decrease with $A$ and therefore the
contribution of the FSI becomes more and more important.
This significant $A$ dependence can be seen in the ratio of these
two curves which is shown in Fig.~\ref{F4}~(bottom).
 We see that the ratio decreases to values about $\simeq 0.4$ for heavy
nuclei. This is also in agreement with what we find for photoproduction where
there is no initial state distortion. 
From this figure we can conclude that in the
$A$-dependence there is indeed valuable information concerning
the $\Lambda(1520)$ absorption and hence, the $\Lambda(1520)$ width in the
medium, which is the main conclusion of the present work.

The total result including all the many-body effects discussed above, but for
other projectile energies with $T_p = 2.7$ and $2.5$~GeV, is shown in
Fig.~\ref{F5}~(top).
Also in Fig.~\ref{F5}~(bottom) we show the total results for $T_p=2.9$~GeV 
but multiplying
artificially the $\Lambda(1520)$ selfenergy by different factors in order to
check the sensitivity of the observable to the value of the
width.
This curves could serve
to get a fair answer about the $\Ls$ width in the medium by
comparing with experimental results.

\section{Summary}
In summary, we have discussed the mass dependence of
the $\gamma$- and $p$-induced production of the $\Lambda(1520)$ hyperons 
from nuclei. The main motivation for this study 
is a prediction of the spectacular increase
of the width of the $\Lambda(1520)$ in the nuclear medium which can reach the
factor $\sim 5$ at normal nuclear matter density. Indications
that this might be the case can be seen in the analysis of $\Lambda(1520)$
production in heavy ion reactions~\cite{R1,R2}.

We have shown
that both reactions, the $\gamma$- and $p$-induced $\Lambda(1520)$ production
can provide an interesting tool to investigate 
the properties
of the $\Lambda(1520)$ hyperon and particularly 
the modification of its  width in the nuclear
medium. The calculations presented here predict a considerable reduction
 of flux of 
the $\Lambda(1520)$ in heavy nuclei with respect to light ones.
We have also shown that the opening of the in-medium absorption channels 
and associated increase of the width of the $\Lambda(1520)$ 
with increasing nuclear matter density is a large source of 
such reduction in the proton induced production and practically 
the only one in the case
of photoproduction. These   effects  are significant, and devoted experiments, 
easily within reach in present facilities like $Spring\, 8$, $ELSA$ and $COSY$, 
can provide good 
information on that magnitude, by measuring the cross sections studied here.

\newpage

\begin{figure*}[t]
\begin{center}
\includegraphics[clip=true,width=0.79\columnwidth,angle=0.]
{Figure5a.eps}
\includegraphics[clip=true,width=0.79\columnwidth,angle=0.]
{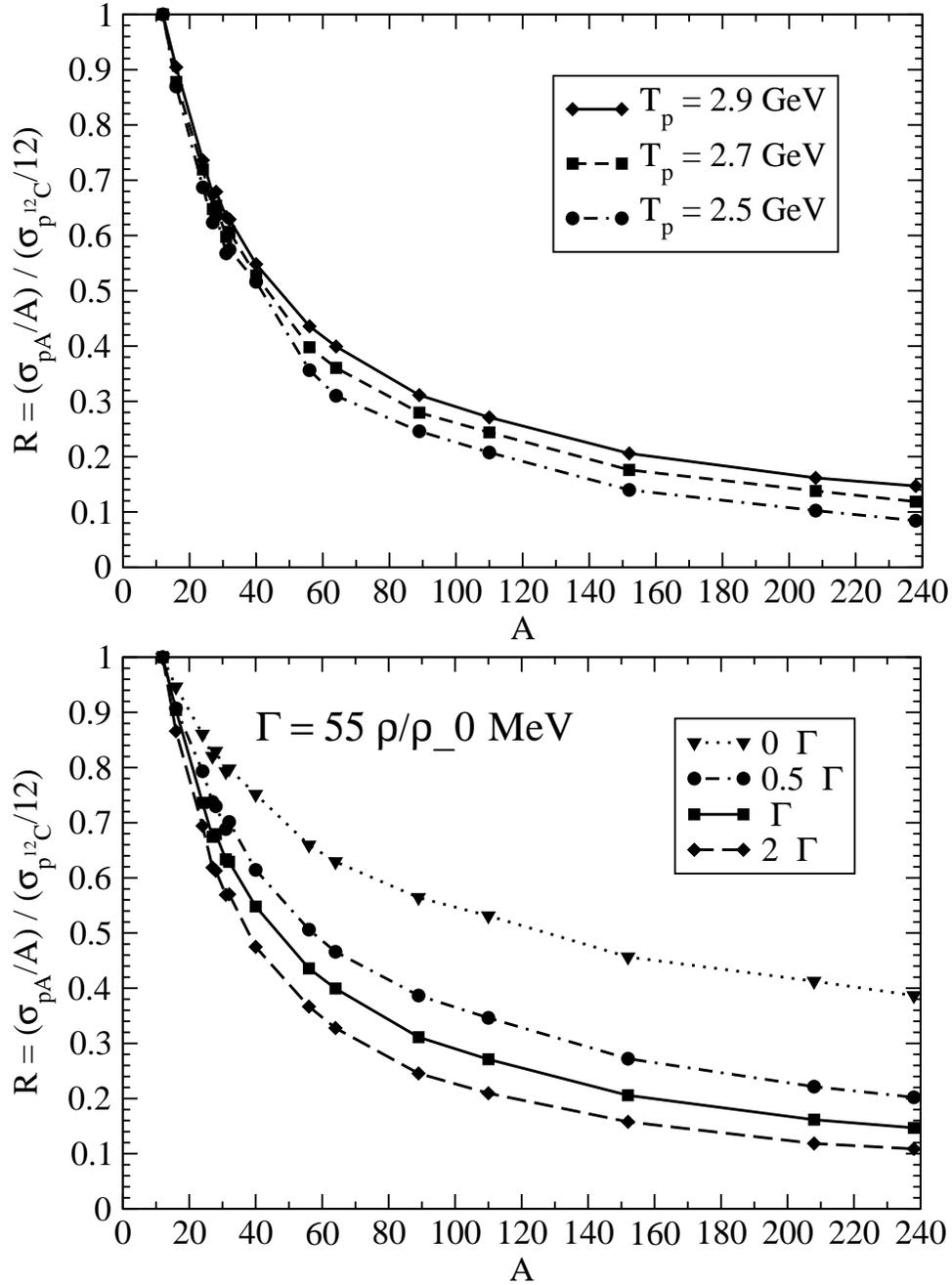}
\caption{\label{F5} \small Top panel:
Ratio of the nuclear cross section normalized to
$^{12}C$ for $T_p=2.5$~(dot-dashed), $2.7$~(dashed) and
$2.9\textrm{ GeV}$~(solid).
Bottom panel: 
Ratio of the nuclear cross section normalized to
$^{12}C$ for $T_p=2.9\textrm{ GeV}$ multiplying the $\Lambda(1520)$ width
in the medium by different factors.}
\end{center}
\end{figure*}

\end{document}